\documentstyle[aps,preprint,eqsecnum]{revtex}
\begin{document}
\input amssym.def
\input amssym
\def\gz{\Bbb Z}
\def\rz{\Bbb R}
\def\half{\mbox{\small$\frac{1}{2}$}}
\def\quart{\mbox{\small$\frac{1}{4}$}}
\def\eighth{\mbox{\small$\frac{1}{8}$}}
\def\threehalf{\mbox{\small$\frac{3}{2}$}}
\def\oneon{\mbox{\small$\frac{1}{N}$}}
\def\be{\begin{equation}}
\def\ee{\end{equation}}
\def\bea{\begin{eqnarray}}
\def\eea{\end{eqnarray}}
\def\coup{K}
\def\G{\frac{\chi^2}{dof}}
\renewcommand{\baselinestretch}{1.}
\baselineskip 3ex
\begin{titlepage}
%
\begin{flushright}
KL-TH-94/5\par
CERN-TH.7183/94
\end{flushright}
\vspace{3ex}
{\LARGE \bf
\centerline{   Three-State Anti-ferromagnetic Potts Model}
\centerline{   in Three Dimensions:} 
\centerline{   Universality and Critical Amplitudes}    
            }
\vspace{0.75 cm}
\begin{center}
{\large    Aloysius P. Gottlob}

\vspace{0.2 cm}
{\it    Universit\"at Kaiserslautern, D-67653 Kaiserslautern, Germany}
\vspace{0.35 cm}

and

\vspace{0.35 cm}
{\large     Martin Hasenbusch}

\vspace{0.2 cm}
{\it    CERN, Theory Division, CH-1211 Gen\`{e}ve 23, Switzerland}
\end{center}
\setcounter{page}{0}
\thispagestyle{empty}
\begin{abstract}\normalsize
We present the results of a Monte Carlo
study of the three-dimensional anti-ferromagnetic
3-state Potts model. We compute various cumulants in the
neighbourhood of the critical coupling. The comparison of the results with
a recent high statistics study of the 3D XY model strongly supports
the hypothesis that
both models belong to the same universality class. From
our numerical data of the anti-ferromagnetic 3-state Potts model
we obtain for the critical coupling
$\coup_c=0.81563(3)$, and for the static critical exponents
$\gamma /\nu=1.973(9)$ and  $\nu=0.664(4)$. \\
\end{abstract}
\nopagebreak
%
\nopagebreak
\begin{flushleft}
\vspace{2 cm}
KL-TH-94/5\newline
CERN-TH.7183/94\\
March 1994
\end{flushleft}
\vspace{1ex}
\end{titlepage}

\newpage
\section{Introduction}

Restoration of symmetries plays a crucial role in lattice field theory.
The Euclidean invariance is broken by the lattice. In order to recover
the continuum field theory, the Euclidean symmetry
has to be restored in the critical
limit of the lattice theory \cite{Lehrbuch}.

Symmetries of the spin-manifold might also be enhanced:
the Hamiltonian of the discrete Gaussian model has only $\gz$ symmetry,
while for sufficiently high temperature
the renormalization group fixed point of the 2D system
is  Gaussian, possessing  $\rz$ invariance \cite{KT}.

In a similar fashion the 3D anti-ferromagnetic (AF) 3-state Potts model is
believed to restore $U(1)$ symmetry at the critical point.
Banavar, Grest and Jasnow \cite{banavar}
 conjectured, based on $\epsilon$-expansion and
Monte Carlo simulations, that the AF 3-state Potts model in three
dimensions undergoes a second-order phase transition and that it shares
the universality class with the $O(2)$ vector model. They expect an analogous result
for the AF 4-state Potts model, sharing
universality with the $O(3)$ vector model.
Hoppe and Hirst \cite{hoppe} concluded from their Monte Carlo simulation
that the AF 5-state Potts model still has a second-order phase
transition, while the AF 6-state Potts model does not
undergo a phase transition at all.

A different point of view is taken by Ono \cite{ono}, who claims to find
a Kosterlitz-Thouless phase transition for the 3D AF
3-state Potts model.
Ueno et al. \cite{ueno} read from their Monte Carlo data that the
3D AF 3-state Potts  model exhibits a
second-order phase transition but with critical exponents different
from the $O(N)$-invariant universality.

Wang et al.  \cite{wang} 
later simulated the 3D AF 3-state Potts model using the cluster algorithm
and found critical exponents in reasonable agreement with  
those of the 3D $XY$ model obtained by field theoretical methods
\cite{wilson72,guillou80,guillou85a}.

In order to clarify this issue we simulated the 3D AF 3-state Potts model
using the single cluster algorithm proposed by Wang et al. \cite{wang},
with  high statistics.

Privman et al. \cite{privman} pointed out that, in addition
 to critical exponents,
universality classes are characterized by critical amplitudes.
In a numerical study,
their values can be obtained more accurately
than those of critical exponents;
hence we determined carefully various cumulants
at the critical point \cite{meyer}. It turned out to be crucial to  use
 an appropriate
order parameter, that allows us 
to detect the restoration of the $U(1)$-symmetry
at criticality.

This paper is organized as follows.
In section 2 we discuss various choices of order parameters and give 
definitions of quantities we are measuring.
In section 3 we discuss our numerical results. We give a comparison of  our
results with previous studies of the 3D AF 3-state Potts model
and 3D $XY$ model in section 4.
Finally section 5 contains our conclusions.

\section{Order parameter and Cumulants}

The $3$-state Potts model
 in three dimensions is defined by the
partition function
   \begin{equation}
   {\cal Z}= \prod_{l\in\Lambda}\sum_{\sigma_l=1}^{3}
   \exp(\coup \sum_{\langle i,j\rangle}
   \delta_{\sigma_i, \sigma_j})\; ,
   \label{a}
   \end{equation}
where the summation is
taken over all nearest neighbour pairs of sites $i$ and $j$  on a
simple cubic lattice $\Lambda$ and
$\coup=\frac{J}{k_b T}$ is
the reduced inverse temperature.
The 3-state Potts model can be transformed into the $\gz$$_3$ clock model
   \begin{equation}
   {\cal Z}= {\cal Z}_0 \prod_{l\in\Lambda}\sum_{\vec{s}_l \in \gz_3}
   \exp(\widetilde{\coup} \sum_{\langle i,j\rangle}
    \vec{s}_i \cdot \vec{s}_j)\; ,
   \label{b}
   \end{equation}
where ${\cal Z}_0 = \exp(-\frac{1}{3} \coup D V) $ with $ V $ being
the volume of the lattice and $\widetilde{\coup} = \frac{2}{3} \coup$.
 In the following we will consider anti-ferromagnetic interactions $J<0$.

For the definition of an order parameter we map the anti-ferromagnetic model
onto a ferromagnetic one. Therefore
we subdivide the simple cubic lattice into  two
sub-lattices $\Lambda^a$ and $\Lambda^b$ in checker-board fashion,
and obtain a positive sign of the coupling
by redefining the spins on one of the sub-lattices
\bea
\vec{s}_i\;' &=& \phantom{-}\vec{s}_i\qquad {\protect\rm for}\quad
i\in\Lambda^a \; ,\nonumber\\
\vec{s}_i\;' &=& -\vec{s}_i\qquad {\protect\rm for}\quad i\in\Lambda^b \; .
\label{Z3}
\eea
Hence we obtain for the magnetization of the ferromagnetic redefined model
\bea
\vec{M}&=&\sum_{i\in\Lambda}\vec{s}_i\;' \nonumber\\
       &=&\sum_{i\in\Lambda^a}\vec{s}_i - \sum_{i\in\Lambda^b}\vec{s}_i\; .
\eea
Note that this definition of the magnetization is the same as the one given
by Ono \cite{ono}. Wang et al. \cite{wang} used a variant choice of the order parameter.
They define
\be
m_\mu=\frac{2}{L^D}\left(\sum_{i\in\Lambda^a}\delta_{\sigma_i,\mu} -
                           \sum_{i\in\Lambda^b}\delta_{\sigma_i,\mu}\right)\; ,
\ee
where $\mu$ takes the values $1,2,3$ and the order parameter is given by
\be
\widetilde{M}=\frac{1}{3}\sum_{\mu=1}^3 |m_\mu|\; .
\ee
It is not obvious how the restoration of the $U(1)$ symmetry can be detected
using this order parameter, hence in the following we only consider $\vec{M}$.

The energy density $E$ of the model is given by
  \be
  E =  \frac{1}{3 L^3} \sum_{\langle i,j \rangle}\langle
       \delta_{\sigma_i, \sigma_j}\rangle .
  \ee

The magnetic susceptibility $\chi$ gives the reaction of the magnetization
to an external field. In the high
temperature phase one gets
  \be
  \chi = \frac{1}{L^3} \langle  \vec{M}^2 \rangle  ,
  \ee
since $ \langle \vec{M} \rangle = 0$. At the critical point the susceptibility
diverges as
\be
\chi \sim L^{\frac{\gamma}{\nu}}\; ,
\label{chifinite}
\ee
where $\nu$ is the critical exponent of the correlation length and $\gamma$
the critical exponent of the susceptibility.

We studied the fourth-order cumulant of the magnetization
\be
U_L= 1-\frac{1}{3}\frac{\langle\vec{M}^4\rangle}{\langle\vec{M}^2\rangle^2}\; .
\ee
In addition we consider the magnetization on sub-blocks of size
$L/2$.
We computed the fourth order cumulants
defined on these sub-blocks and a normalized nearest-neighbour product
\be
N\!N= \frac{\langle\vec{M}_I \vec{M}_J\rangle}{\langle\vec{M}_I^2\rangle}\; ,
\ee
where $I$ and $J$ are nearest-neighbour-blocks.
At the critical point the cumulants should converge to a universal fixed
point. This property is used to determine the critical coupling \cite{binder}.

\section{Numerical results}

In the present work we employ the single cluster algorithm proposed by Wang
et al. \cite{wang}. In the ferromagnetic
$\gz$$_3$ parametrization of eq.(\ref{Z3}) this algorithm can be understood
as the single cluster algorithm introduced
by Wolff \cite{wolff89a}, with the reflection planes  restricted by the
$\gz$$_3$ symmetry.

On lattices of the size $L =  4,8,16,32 $ and 64 we have performed simulations
at $K_0=0.8157$, which is the estimate for the critical coupling
obtained in ref. \cite{wang}.
We performed $N$ measurements taken every $N_0$
update steps.  We have chosen $N_0$ such  that on  average
the lattice is updated approximately twice for one  measurement.
 The results of the runs are
summarized in Table\ \ref{crres1}.

\subsection{Critical coupling}

First we determined the critical coupling $K_c$,
employing Binder's phenomenological renormalization
group method \cite{binder}.

For the extrapolation of the observables, entering the cumulants, to couplings
$K$ other than the simulation coupling $K_0$,
we used the reweighting formula \cite{swendferr}
  \begin{equation}
   \langle A \rangle (K) = \frac{\sum_i A_i \exp((-K+K_0) H_i)}
                     {\sum_i  \exp((-K+K_0) H_i)} ,
  \label{reweight}
  \end{equation}
where $i$ labels the configurations generated according to the Boltzmann-weight
at $K_0$. We computed the statistical errors from Jackknife binning
\cite{siam} on the
final result of the extrapolated cumulants. The extrapolation gives good
results only within a small neighbourhood of the simulation coupling $K_0$.
This range shrinks with increasing volume of the lattice.
However,
the figs. \ref{rewcum} a)-c) show that
the extrapolation performs well in a sufficiently large neighbourhood
of the crossings of the cumulants.

When one considers the
cumulants as functions  of the coupling, the crossings of the curves for
different $L$
provide  an estimate for the critical coupling $K_c$.
The results for the crossings are summarized in Table\ \ref{crosscoup}.

The convergence of the crossings of $N\!N$ towards $K_c$ seems to be slower
than that of the fourth-order cumulants, but
it is interesting to note that the $K_{cross}$
for the fourth-order cumulant and
$N\!N$  come from different  sides with increasing  $L$.
This is shown in fig.\ \ref{kccross}, where the estimates of $K_{cross}$
versus the lattice size $L$ are plotted.
This behaviour
we also observed for the 3D XY model \cite{wexy}.
The given errors are calculated with a jackknife procedure.
The convergence of the crossing coupling $K_{cross}$ towards $K_c$ should
follow
  \be
  K_{cross}(L)  =  K_c \;( 1 + const.  L^{ -(\omega+\frac{1}{\nu})}+\ldots),
  \label{kcross}
  \ee
where $\omega$ is the  correction to scaling exponent \cite{wegner,binder}.
We performed a two-parameter fit for the crossings of the cumulants,
keeping the exponents fixed to $\omega=0.780$ and $\nu=0.669$ \cite{guillou85a},
following the above formula. It is important to note that the value of
$K_c$ does not depend strongly on the value of the exponents.
Our final estimate for the critical coupling
is $K_c = 0.81563(3)$, obtained from the two-parameter fit eq.(\ref{kcross})
of the fourth-order  cumulants.
For the cumulant on the sub-block
we have to discard the $4-8$ crossing to obtain a fit with an acceptable
$\chi^2/dof$.
The results of the fits are summarized in Table\ \ref{kcfit}.

\subsection{Critical amplitudes}

At the critical coupling
$K_c$ the cumulants converge with increasing lattice size $L$
 to an universal fixed point.
 The results of the  cumulants at $K=0.81560, 0.81563$ and $0.81566$,
which are our best estimate of the critical coupling and the edges of
the error-bar,
 are given in Table\ \ref{renobs}.

The convergence rate is given by \cite{binder}
  \be
  U_{L}(K_c)  =  U_{\infty} \;( 1 + const.  L^{ -\omega}+\ldots) \, .
  \label{ufix}
  \ee

We performed a two-parameter fit with $\omega=0.780$ \cite{guillou85a}
being fixed. The results of the fits are given in Table\ \ref{cumfit}.
We had to discard the data from $L=4$ in order to obtain an acceptable
$\chi^2/dof$. The results of the fits are stable when discarding the $L=8$
data point.

As our final estimates we obtain\hfill\newline
\begin{center}
\begin{tabular}{lcccc}
\multicolumn{5}{c}{Final estimates}\\
Model  & $K_c$   &   $U_L$    &   $U_{L/2}$   &  $N\!N$ \\
Potts  & $0.81563(3)$ & $0.5861(6)$  & $0.5465(6)$  & $0.8123(9)$\\
$XY$   & $0.45419(2)$ & $0.5875(28)$ & $0.5477(25)$ & $0.8144(37)$
\end{tabular}
\end{center}
where we have taken into account the uncertainty of the critical
coupling $\coup_c$.
For comparison we give the analogous results for the 3D XY model.

A careful reanalysis, taking into account eq.(\ref{kcross}), leads
to a small shift in $K_c$ compared to ref. \cite{wexy} where we quoted
$K_c=0.45420(2)$ as final result. This shift of the coupling also
implies a slight change in the result for the cumulants. The higher
accuracy of the values for the Potts model is  due to 
higher statistics and to a smaller variation of the fit results
within the error-bars of the critical coupling.

Note that the accurately determined values of the cumulants
of the two models coincide
within the error-bars. This fact strongly favours the hypothesis that the 
two models belong
to the same universality class.

\subsection{Critical exponents}
We extracted the critical exponent  $\nu$ of the correlation length from the
$L$ dependence
of the slope of the fourth-order cumulants and the nearest-neighbour
observable at the estimated critical coupling \cite{binder}.
According to Binder, the scaling relation for the slope of the fourth-order
cumulant is given by
  \be
  \left .\frac{\partial U(L,\coup)}{\partial\coup} \right |_{\coup_c}  \propto
  L^{1/\nu}.
  \label{slope}
  \ee
We evaluated the slopes of the observables $A$ entering  the cumulant $U$
according to
  \be
  \frac{\partial \langle A \rangle }{\partial \coup} = \langle A H \rangle -
  \langle A \rangle \langle H \rangle ,
  \ee
where $A$ is an observable and $H$ is the energy.
The statistical errors are calculated from  a Jackknife analysis for the value
of the slope. First we estimated the exponent $\nu$ from different lattices via
  \be
  \nu = \frac{\ln \left( L_2 \right)- \ln\left(L_1\right)}{\displaystyle
  \ln \left(
  \left.\frac{\partial A(L_2,\coup)}{\partial\coup} \right |_{\coup_c}\right)-
  \ln \left(\left.\frac{\partial A(L_1,\coup)}{\partial\coup} \right
  |_{\coup_c}\right)}.
  \label{nulat}
  \ee
The results are given in Table\ \ref{nutab}.
Since the statistical errors are small one can check the convergence
of $\nu$ obtained from the pair of lattice sizes $L$ and $2L$
with increasing $L$. The value of $\nu$ obtained from the
fourth-order cumulant at $L=16$ coincides
within the error-bars with the  $L=32$ result. The convergence of the
$\nu$ computed from $N\!N$ is worse. The $L=16$ and $L=32$ results are
off by more than three times the error-bars.
We also performed a two parameter fit to eq.(\ref{slope}).
The estimates of the fits are given in Table\ \ref{nufittab}.
 When we discard
the data stemming from the  $L=4$ and 8 lattices, we get a $\G$ smaller then 1.
If we take the uncertainty of the critical coupling into account
we obtain the following values
for the exponent $\nu$:\newline
\begin{center}
\begin{tabular}{ccc}
\multicolumn{3}{c}{Estimates of $\nu$}\\
  $U_L$        & $U_{L/2}$    &  $N\!N$ \\
 $0.6639(38)$  & $0.6592(23)$ & $0.6924(23)$\\
\end{tabular}
\end{center}
One has to note that a small $\G$ does not mean that errors due to corrections
to scaling are negligible. We have seen in the discussion of the cumulants
that $N\!N$ is much more affected by corrections to scaling than the
fourth-order cumulant.
Hence the result for $\nu$ obtained from $N\!N$ should
not be
taken  too seriously.   As our final result, we take the value $\nu=0.664(4)$
obtained from
the fourth-order cumulant on the full lattice.

In order to estimate the ratio $\gamma/\nu$ of the critical exponents we
studied
the scaling behaviour of the magnetic susceptibility
defined on the  full lattice and on sub-blocks.

The dependence of the susceptibility on the lattice size
at the critical point is given by eq.(\ref{chifinite}).
We have estimated $\gamma/\nu$ from pairs of lattices with size $L_1$,
$L_2$. The ratio then is given by
  \be
  \frac{\gamma}{\nu} = \frac{\ln(\chi(L_1,\coup_c))-\ln(\chi(L_2,\coup_c))}
  {\ln(L_1)-\ln(L_2)} .  \label{phen}
  \ee
Table\ \ref{gamnu} shows the estimates of the ratio.
The estimates for $\gamma / \nu$ obtained
from the sub-blocks increase
with increasing lattice size $L$, while those obtained from the full lattice
decrease. We also  performed a two-parameter fit following
eq.(\ref{chifinite}).
We could not extract reliable estimates for the ratio from the fits. However,
the estimates of $\gamma/\nu$ from $\chi_L$ and $\chi_{L/2}$ obtained
from the largest lattices coincide within the error-bars.
Hence we take
$\gamma / \nu=1.973(9)$ obtained from the ratio of $\chi_L$ of the largest
lattices as our final result, where
statistical  as well as systematic errors should be covered.
 Using the scaling relation
$\eta = 2 - \frac{\gamma}{\nu}$,
we obtain for the anomalous dimension $\eta=0.027(9)$.

\subsection{Symmetry of the order parameter}

In a qualitative fashion we also looked at the symmetry of the order parameter
$\vec{M}$. In figs.\ \ref{ordfig} a)$-$c) we have plotted the probability
distribution of $\vec{M}$ at $\coup= 2.0$ on lattices with $L=8,16$ and $32$.
Fig.\ \ref{ordfig} b) shows that
the probability distribution for $L=16$ is
  strongly peaked at six locations. Three
of these peaks are considerably larger than the other ones.

This can be understood as follows. Configurations of minimal energy
are reached when
on one sub-lattice all spins take the same value while on the other sub-lattice
the spins can take any of the other two values \cite{ono}.
The cluster algorithm easily manages to change the value of the spin on
the ordered sub-lattice, while it takes many updates to switch the order from
one sub-lattice to the other. For increasing $L$ the peaks become more
pronounced and the tunnelling times between the two metastable states increase.

In c) ($L=32$) the simulation
time (we have plotted 5000 measurements of the order parameter at $K=2.0$) was
not large enough to see a flip of the ordered state from one sub-lattice
to the other, while for $L=4$ the tunnelling time is much smaller than the
simulation time.

Figs.\ \ref{ordfig} d)$-$e) show the probability distribution
of the order parameter on lattices with $L=8,16$ and $32$
at $K=0.8157$ near the estimate of the critical coupling.
In contrast to the situation at low temperature $K=2.0$ we could not observe
deviations from $U(1)$ invariance at the critical point on any of the
considered lattices.
Thus the Potts model seems to restore $U(1)$ symmetry at the critical point.

For short range interactions a universality class should be characterized
by the dimensionality of the system
and the symmetry of the order parameter at criticality
\cite{kadanoff}.
Hence we conclude from the distribution of the order parameter at the
critical point of the 3-state Potts model,
 that this model shares the universality
 class with the XY model.

\subsection{Performance of the Algorithm}

The efficiency of a stochastic algorithm is characterized by the
integrated autocorrelation time
   \begin{equation}
   \tau_{int} ={1 \over 2} \sum_{t = - \infty}^{\infty} \rho (t) \, ,
   \label{f}
   \end{equation}
where  the normalized autocorrelation function $\rho (t)$ of an observable
$A$ is given by
   \begin{equation}
   \rho (t) = {{ \langle A_i \cdot A_{i+t} \rangle - {\langle A \rangle}^2}
   \over{\langle A^2 \rangle - {\langle A \rangle}^2 }} \,  .
   \label{e}
   \end{equation}
We  calculated the integrated autocorrelation times
$\tau_{int}$ with a self-consistent truncation window of width 6$\tau_{int}$
for the energy density $E$ and the magnetic susceptibility $\chi$ for lattices
with $L = 4$ up to $L=64$ at the  coupling $\coup=0.8157$.
Our estimates for the critical dynamical exponents are $z_{E} = 0.18(3)$ and
$z_{\chi}=0.09(3)$ taking only statistical errors into account.
Note that these exponents
are consistent with those found in cluster simulations of the 3D XY model
 \cite{janke90a,wexy}.

Finally let us briefly comment on the CPU time.
160 single cluster updates of the $64^3$ lattice at the  coupling
$K = 0.8157$ plus one measurement of
the observables took on average 20 sec CPU time on a IBM RISC 6000-550
workstation. For comparison, 160 single cluster updates of the $64^3$ lattice
at the coupling $K=0.45417$ of the 3D $XY$ model took on average 26 sec
CPU time on the same machine \cite{wexy}.
All our MC simulations of the 3D AF Potts model together
took about one month of CPU-time on an IBM RISC 6000-590 workstation where the
simulations were done.

\section{Discussion of the  results}
In this section we compare our results with those obtained in previous studies
of the 3D AF 3-state Potts model. Furthermore we compare the AF 3-state
Potts results with critical exponents and amplitudes obtained for 3D
O(N)-vector
models.
In Table\ \ref{compare} we display estimates for the critical properties of the
3D AF 3-state Potts model and the 3D $XY$ model.

Our value for $K_c$ agrees with the value obtained from the MC study
by Wang et al. \cite{wang} within the error-bars.
However, our error estimate is about 15 times more accurate.
The Monte Carlo result for $K_c$ given by Hoppe and Hirst \cite{hoppe}
is  consistent with our value within twice  their error estimate, while
our error estimate is about 84 times smaller than the error they quote.
Yasumura et al. \cite{yasumu}
obtained from the high temperature series expansion
a result for $K_c$ that
is about 350 times our error-bar larger
than our value. One has to note that they did not extract an estimate
of the error.

The estimate for the ratio $\gamma/\nu$ that we obtain is consistent
with the
one given by Wang
et al. \cite{wang}. Even taking into account systematic
errors, we could reduce the uncertainty by about a factor of 3.
Ueno et al. \cite{ueno} calculated the ratio $\gamma/\nu$ via an interface
approach. Their estimate is about 40 times their error-bar smaller than our
value, while our error is about 2 times smaller than the one they qoute.
 Hence we can
rule out their result with high confidence.

The ratio $\gamma/\nu$ obtained from a MC study of the 3D $XY$ model
by the authors \cite{wexy} is in excellent agreement
with the one obtained in this work for the 3D Potts model.
Our value for the ratio  $\gamma/\nu$ is  also
consistent within the error-bars with the
value obtained from resummed
perturbation series of the $2$-component $\phi^4$ theory in 3D by Guillou
and Zinn-Justin \cite{guillou80}.

Our estimate of $\nu$ coincides within the error-bar
with the value of the MC study
of Wang et al. \cite{wang} while our error is about 6 times smaller than their
error estimate. The value for $\nu$ given by Ueno et al. \cite{ueno} is about
8 times their error-bar smaller than our value, while our error
estimate is about
3 times smaller.

Our estimate for the exponent $\nu$ agrees within the error-bars
 with the one
obtain for  the 3D $XY$ model in a MC study by the authors \cite{wexy}.
Moreover it is consistent
within the quoted error-bars  with the estimate obtained by
 resummed perturbation series
of the $2$-component $\phi^4$ theory in 3D by Guillou and Zinn-Justin
\cite{guillou80}.

The most accurate value of $\nu$ is obtained in a $^4$He experiment by
Ahlers and Goldner \cite{ahlers}.
Our estimate of $\nu$ is consistent
with the experimental estimate of $\nu$ for $^4$He
within two times our error estimate. The experimental value is about 3
times more accurate than our result.

In order to judge  the relevance of this nice agreement of the exponent
$\nu$ of the 3D  AF 3-state Potts model and the 3D 2-component
 $\phi^4$ theory, one should note that Landau and Ferrenberg obtained
\cite{landau} $\nu=0.6289(8)$ in a Monte Carlo study of the 3D Ising model,
which is consistent with the resummed perturbation theory result
for the  1-component $\phi^4$ theory $\nu=0.6300(15)$ \cite{guillou80}.

The value $\nu=0.704(6)$ obtained in a Monte Carlo study of the 3D $O(3)$
model by Janke and Holm \cite{jankeo3},
which can be compared with the 3-component $\phi^4$ theory
result $\nu =0.705(3)$ \cite{guillou80}  is also
clearly inconsistent with the AF
3-state Potts value.

There exists no previous published result for the fourth-order cumulant of the
order parameter $\vec{M}$.
Hence we have to restrict our comparison to
results for the 3D $XY$ model and the 2-component $\phi^4$ theory.  Our estimate
coincides with the one for the 3D $XY$ model \cite{wexy} within the error-bars.
But the value obtained from the $\epsilon$-expansion of the 2-component
$\phi^4$ theory
of Br\'ezin and Zinn-Justin \cite{brezin85a} is off by 57 times our error
estimate.
Note that the relative error of the fourth-order cumulant obtained in this
work is of order 0.1\%
taking into account the uncertainty of the critical coupling.

The results obtained in Monte Carlo studies of the 3D Ising model
by Landau and Ferrenberg \cite{landau} $U_{\infty} \approx 0.47$,
and one of the authors \cite{hasenbus} $U_{\infty} = 0.464(2)$  are clearly
off from our result $U_{\infty} = 0.5862(6)$ for the 3D AF 3-state Potts
model.
The fourth-order cumulant of the 3D  O(3) Heisenberg model, where the
best value $U_{\infty}=0.6217(8)$
stems from Monte Carlo simulation
\cite{jankeo3} is also far off from the 3D AF 3-state Potts result.

\section{Conclusions}

In the present work we have applied the single cluster algorithm
\cite{wolff89a} for the simulation of  the 3D AF Potts model.
 The analysis of the critical
dynamical
behaviour shows that the algorithm is almost free of critical slowing down.
Thus we were able to increase the statistics considerably by extensive use of
modern RISC workstations.

The phenomenological RG approach allowed us to determine the critical
amplitudes of the model with an accuracy of about 0.1\%. From the fit
of the crossings of the fourth-order cumulant we obtain for the
critical coupling of the model $K_c=0.81563(3)$, which reduces the error of
an earlier MC study by a factor of about 15.

The excellent agreement of the universal critical amplitudes and exponents
of the 3D AF Potts model with the ones of the 3D XY model strongly
favours the supposition
that the two models belong to the same universality class.

\acknowledgements

We would like to thank S.~Meyer for stimulating discussions on the subject,
and A.~Kavalov for a careful reading of the manuscript.
The numerical simulations were performed on an IBM RISC 6000 cluster of the
Regionales Hochschulrechenzentrum Kaiserslautern (RHRK). One of us (A.G.)
expresses his gratitude for
the hospitality he enjoyed during a visit to CERN.



\newpage
%
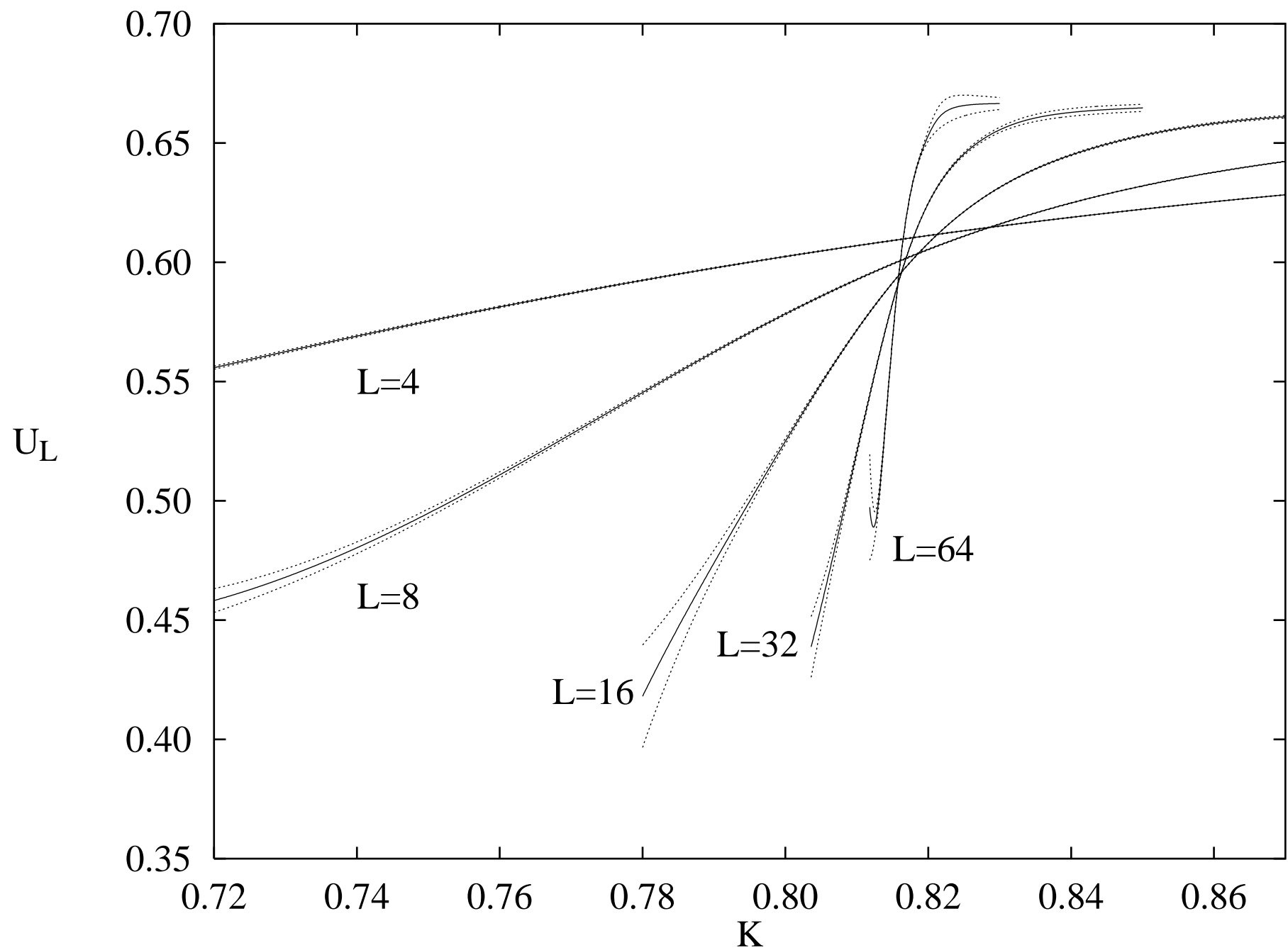
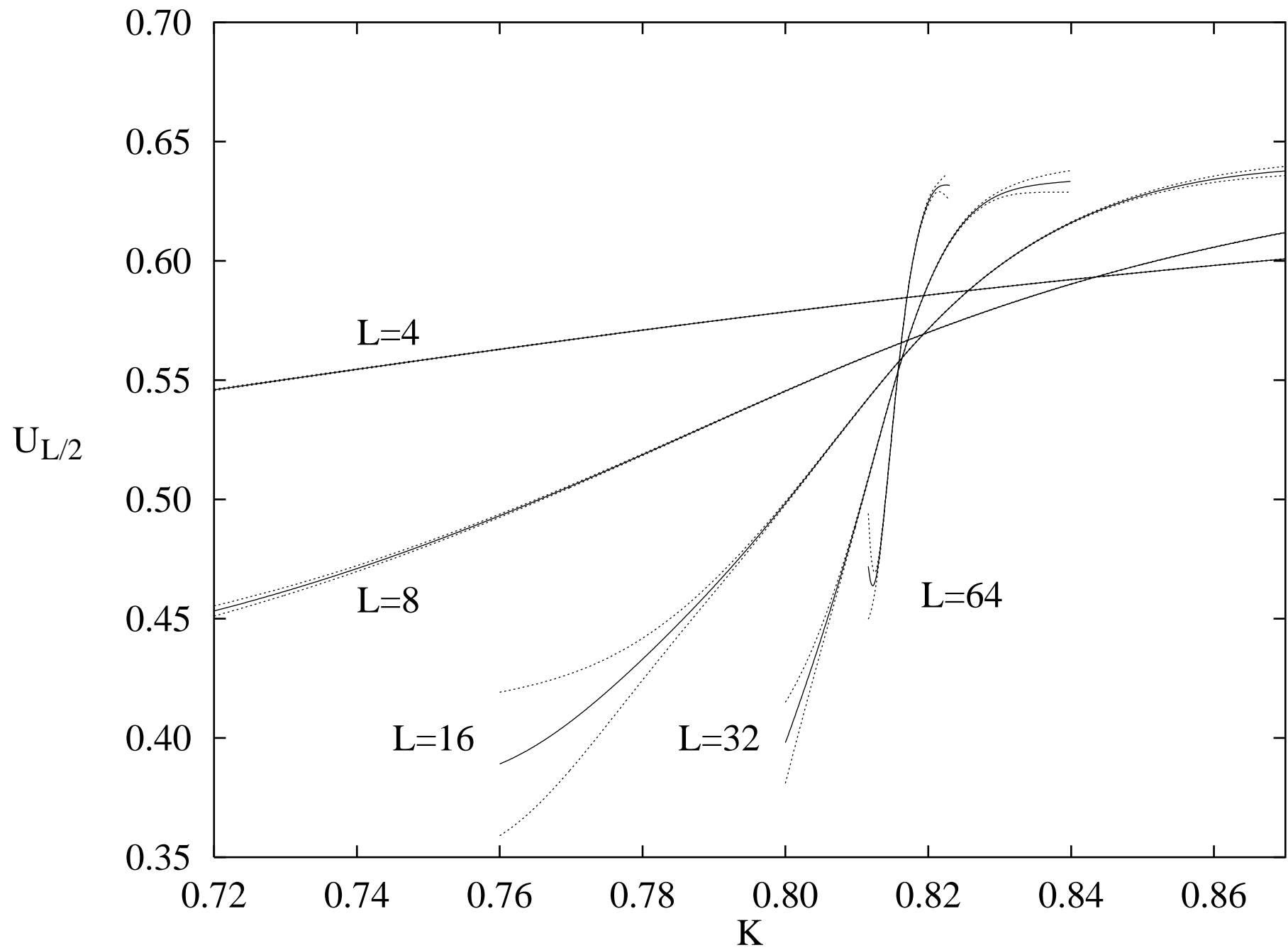
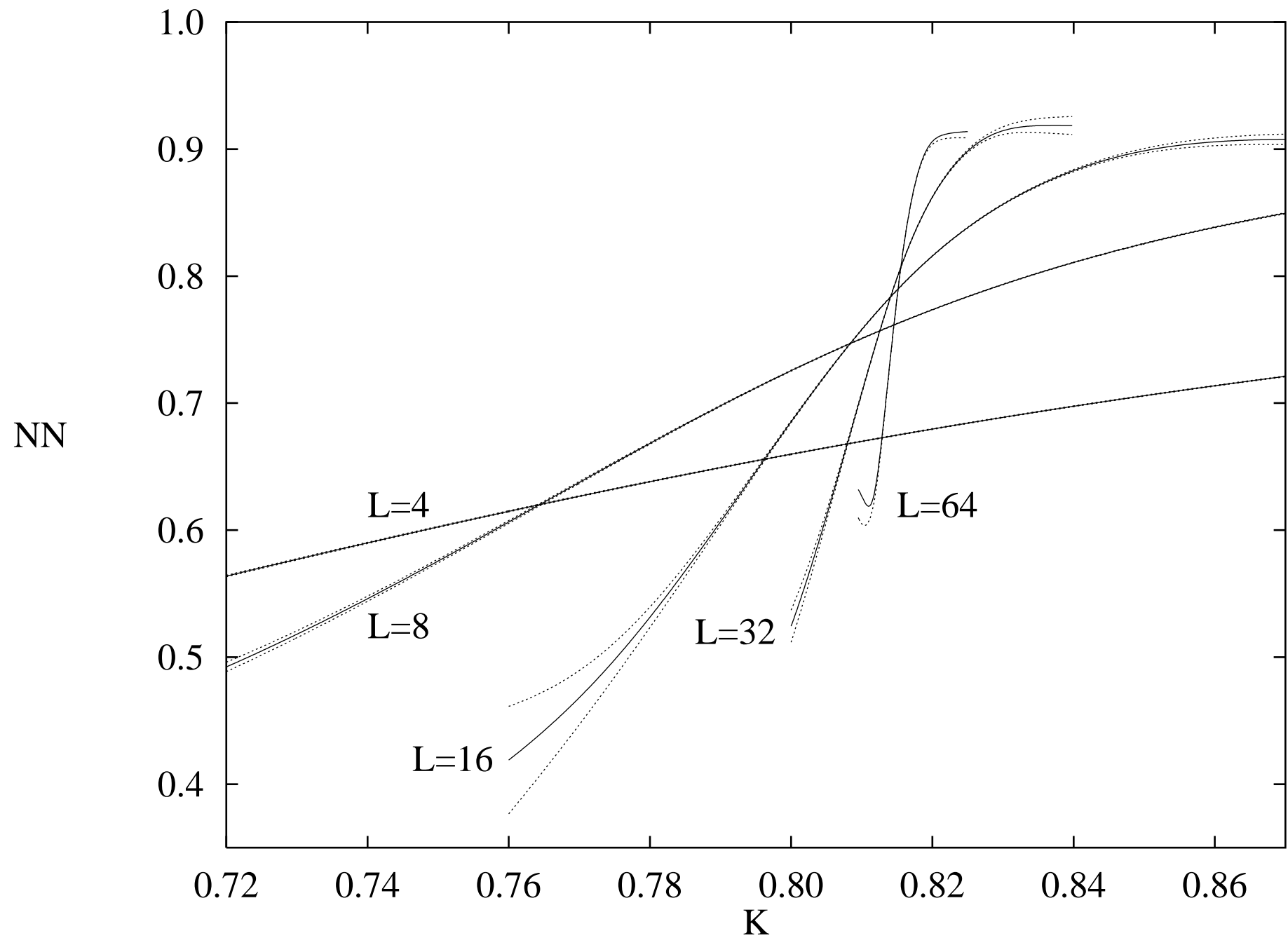
\begin{figure}
\caption{
      Reweighting plots
      from the simulations at $\coup_0=0.8157$. The dotted lines give
      the statistical error obtained by a Jackknife analysis.
      a) shows the cumulant defined on the  full lattice, b) shows
      the reweighting of the cumulant defined on half of the
      lattice and c) shows the reweighting of the nearest neighbour
      product $N\!N$.
\label{rewcum}
        }
\end{figure}
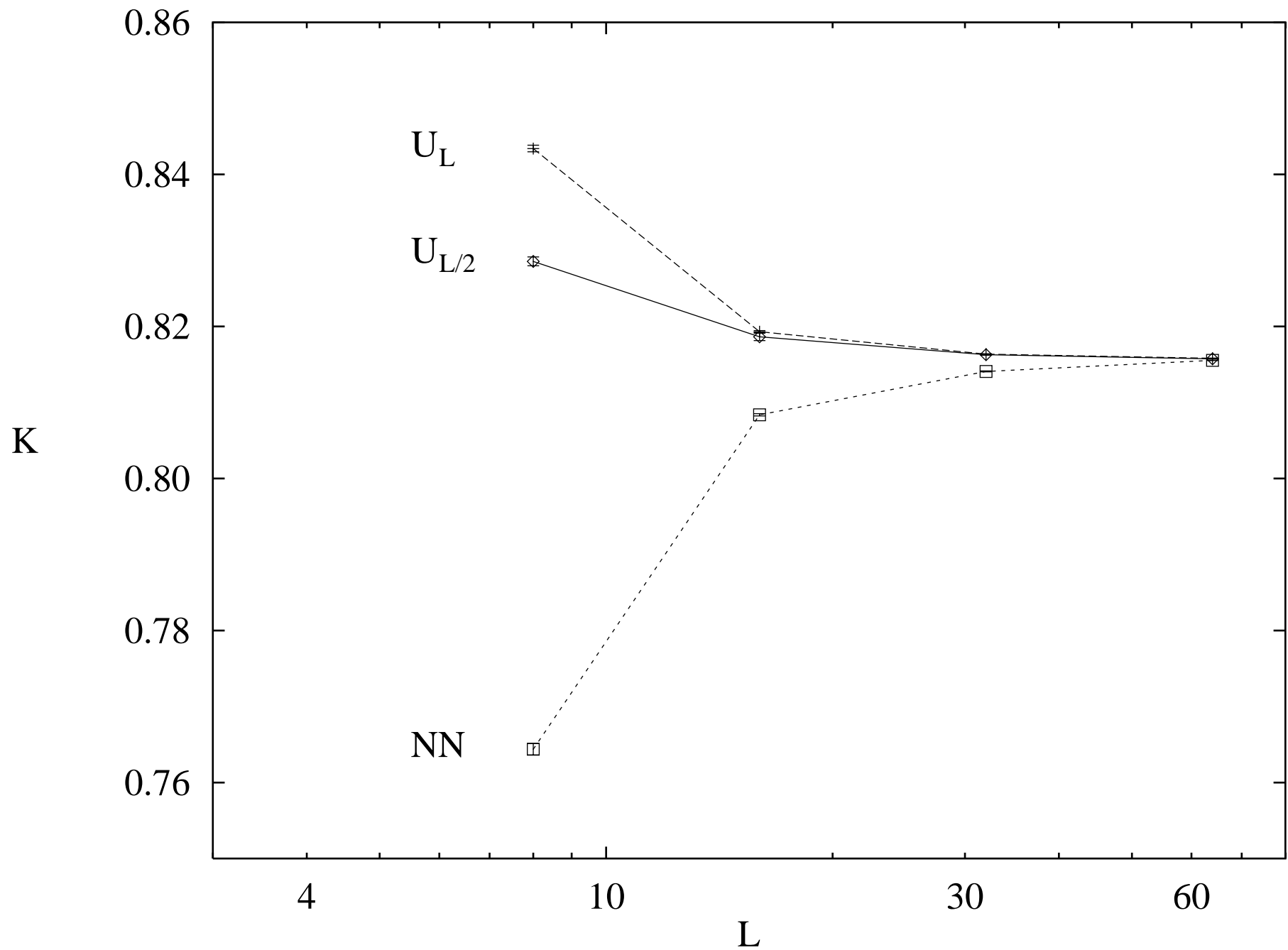
\begin{figure}
\caption{
      Plot of the crossing couplings of the reweighted cumulants.
      Since the statistical error of the reweighted cumulants is
      small, one is able to see systematic convergence of the crossing
      couplings towards the critical coupling $K_c$.
\label{kccross}
        }
\end{figure}


\begin{figure}
\caption{
      Plot of the probability distribution of the order parameter $\vec{M}$.
      The figures a) to c) show the distribution for $K=2.0$ on lattices
      with $L=8,16$ and $32$. The figures from d) to f) show the probability
      distribution for $K=0.8157$ near the final estimate of the critical
      coupling on lattices with $L=8,16$ and $32$.
\label{ordfig}
        }
\end{figure}

%
\begin{table}
\caption{
         Results of the energy density $E$, the susceptibility defined on the
         full lattice ($\chi_L$) and defined on the sub-blocks of size $L/2$
         ($\chi_{L/2}$).
         The data is obtained from  simulations at the fixed
         coupling $\coup_0 = 0.8157$ near the final estimate of the critical
         coupling.
         $\tau$ denotes the integrated autocorrelation time of the specified
         observable, given in units of the average number of clusters that is
         needed to cover the volume of the lattice. The statistics are given
         in terms of $N$ measurements taken every $N_0$ update steps.
         }
         \label{crres1}
\begin{tabular}{rlrlllll}
$L$& $N$ &$N_0$& $E$ & $\tau_E$ & $\chi_L$ & $\tau_{\chi}$ & $\chi_{L/2}$ \\\hline
 4 & 200k&  10 & 0.1064(1)  & 0.55(1)&  16.59(2) &0.54(1) &    21.51(1)\\
 8 & 200k&  20 & 0.12383(4) & 0.61(1)&  69.54(7) &0.56(1) &   80.53(6) \\
16 & 200k&  40 & 0.13006(2) & 0.64(1)& 281.01(33)&0.57(1) &  314.34(28)\\
32 & 200k&  80 & 0.132267(7)& 0.71(1)&1120.8(1.3)&0.60(1) & 1239.7(1.1)\\
64 & 170k& 160 & 0.133062(3)& 0.82(2)&4447.8(5.9)&0.64(2) & 4890.2(4.9)
\end{tabular}
\end{table}

%
\begin{table}
\caption{Estimates of the couplings  at the crossings of the
         the reweighted cumulants. $L_1-L_2$ denotes
         the lattices used to determine the crossing.
        }
         \label{crosscoup}
\begin{tabular}{clll}
\multicolumn{4}{c}{Crossing couplings $K_{cross}$}\\ \hline
$L_1-L_2$& $K_{cross}(U_L)$& $K_{cross}(U_{L/2})$& $K_{cross}(N\!N)$\\ \hline
 4-8   & 0.82941(46)  & 0.84403(33)  & 0.76580(50)  \\
 8-16  & 0.81844(15)  & 0.81916(10)  & 0.80837(13)  \\
 16-32 & 0.816202(53) & 0.816345(43) & 0.814728(41) \\
 32-64 & 0.815737(20) & 0.815795(15) & 0.815529(14) \\
\end{tabular}
\end{table}

%
\begin{table}
\caption{
         Estimates of the critical couplings from the fit following
         eq.(\protect\ref{kcross})
         with $\nu=0.669$ and $\omega=0.78$ being fixed. \# gives the
         number of discarded data points with small $L$.
        }
         \label{kcfit}
\begin{tabular}{cllr}
\multicolumn{4}{c}{ $K_c$ from $U_{L}$}\\ \hline
\#&   $K_c$        &  $C_{U_L}$     &   $\G$  \\ \hline
 0&   0.815619(20) &  0.403(12) &   0.006 \\
 1&   0.815619(22) &  0.404(21) &   0.011 \\ \hline\hline
\multicolumn{4}{c}{ $K_c$ from $U_{L/2}$}\\ \hline
\#&   $K_c$        &  $C_{U_{L/2}}$ &  $\G$   \\ \hline
 0&   0.815530(15) &  0.729(8)  & 185.0   \\
 1&   0.815647(16) &  0.499(14) &   0.198 \\ \hline\hline
\multicolumn{4}{c}{ $K_c$ from $N\!N$}\\ \hline
\#& $K_c$          & $C_{N\!N}$    & $\G$ \\ \hline
 0&   0.815987(14) &$-$1.244(11)   & 263 \\
 1&   0.815854(15) &$-$0.964(17)   & 76.9
\end{tabular}
\end{table}

%
\begin{table}
\caption{Estimates of the cumulants
         at our best estimate of the critical coupling $K_c=0.81563(3)$.}
        \label{renobs}
\begin{tabular}{clllll}
\multicolumn{4}{c}{Estimates of $U_L$}\\ \hline
$L$&  $K_c-\Delta K_c$ & $K_c$ & $K_c+\Delta K_c$ \\ \hline
 4 & 0.60975(21) & 0.60976(21) & 0.60977(21) \\
 8 & 0.59985(25) & 0.59989(25) & 0.59993(24) \\
16 & 0.59367(25) & 0.59378(25) & 0.59388(25) \\
32 & 0.58981(26) & 0.59012(26) & 0.59042(26) \\
64 & 0.58724(28) & 0.58812(28) & 0.58899(27) \\
 \hline\hline
\multicolumn{4}{c}{Estimates of $U_{L/2}$}\\ \hline
$L$&  $K_c-\Delta K_c$ & $K_c$ & $K_c+\Delta K_c$ \\ \hline
 4 & 0.58449(11) & 0.58450(11) & 0.58451(11) \\
 8 & 0.56479(15) & 0.56482(15) & 0.56486(15) \\
16 & 0.55700(18) & 0.55711(18) & 0.55721(18) \\
32 & 0.55223(20) & 0.55253(20) & 0.55283(20) \\
64 & 0.54863(22) & 0.54949(21) & 0.55035(21) \\
 \hline\hline
\multicolumn{4}{c}{Estimates of $N\!N$}\\ \hline
$L$&  $K_c-\Delta K_c$ & $K_c$ & $K_c+\Delta K_c$ \\ \hline
 4 & 0.67621(36) & 0.67624(36) & 0.67627(36) \\
 8 & 0.76378(32) & 0.76385(32) & 0.76392(32) \\
16 & 0.79269(29) & 0.79286(29) & 0.79303(29) \\
32 & 0.80158(28) & 0.80205(28) & 0.80251(28) \\
64 & 0.80351(30) & 0.80479(30) & 0.80606(30)
\end{tabular}
\end{table}

%
\begin{table}
\caption{
         Estimates of the cumulants from the fit following
         eq.(\protect\ref{ufix}). \# gives the
         number of discarded data-points with small $L$.
        }
         \label{cumfit}
\begin{tabular}{cllcllcllc}
\multicolumn{10}{c}{Estimates of $U_L$}\\ \hline
&\multicolumn{3}{c}{ $K_c-\Delta K_c$}&
\multicolumn{3}{c}{ $K_c$}&
\multicolumn{3}{c}{ $K_c+\Delta K_c$}\\ \hline
\#& $U_L$       & $C_{U_L}$   & $\G$ & $U_L$       & $C_{U_L}$
& $\G$
& $U_L$       & $C_{U_L}$   & $\G$  \\ \hline
0& 0.58486(20) & 0.1264(16) &  2.64 &
   0.58541(20) & 0.1231(16) &  0.53 &
   0.58595(20) & 0.1198(16) &  0.87 \\
1& 0.58511(22) & 0.1271(33) &  7.31 &
   0.58560(22) & 0.1227(33) &  2.43 &
   0.58608(21) & 0.1184(32) &  1.14 \\
2& 0.58567(27) & 0.1165(66) & 10.07 &
   0.58605(27) & 0.1128(66) &  3.01 &
   0.58643(27) & 0.1089(66) &  0.53 \\
\hline\hline
\multicolumn{10}{c}{Estimates of $U_{L/2}$}\\ \hline
&\multicolumn{3}{c}{ $K_c-\Delta K_c$}&
\multicolumn{3}{c}{ $K_c$}&
\multicolumn{3}{c}{ $K_c+\Delta K_c$}\\ \hline
\#& $U_{L/2}$       & $C_{U_{L/2}}$   & $\G$ & $U_{L/2}$
& $C_{U_{L/2}}$
& $\G$ & $U_{L/2}$       & $C_{U_{L/2}}$
& $\G$  \\ \hline
0& 0.54308(14) & 0.2218(11) &   87.50 &
   0.54360(14) & 0.2184(11) &  104 &
   0.54410(14) & 0.2152(11) &  125 \\
1& 0.54580(16) & 0.1764(24) &   10.84 &
   0.54625(16) & 0.1721(24) &    2.82 &
   0.54673(16) & 0.1676(24) &    0.09 \\
2& 0.54608(20) & 0.1729(53) &   14.34 &
   0.54643(20) & 0.1695(53) &    3.83 &
   0.54682(20) & 0.1650(53) &    0.01 \\
\hline\hline
\multicolumn{10}{c}{Estimates of $N\!N$ }\\ \hline
&\multicolumn{3}{c}{ $K_c-\Delta K_c$}&
\multicolumn{3}{c}{ $K_c$}&
\multicolumn{3}{c}{ $K_c+\Delta K_c$}\\ \hline
\#& $N\!N$       & $C_{N\!N}$   & $\G$ & $N\!N$  & $C_{N\!N}$
& $\G$ & $N\!N$       & $C_{N\!N}$   & $\G$  \\ \hline
0& 0.83152(23) &-0.5034(15) & 2048 &
   0.83240(23) &-0.5068(15) & 1949 &
   0.83327(23) &-0.5102(15) & 1855 \\
1& 0.81677(25) &-0.3083(28) &  187 &
   0.81751(25) &-0.3127(28) &  170 &
   0.81825(25) &-0.3171(28) &  158 \\
2& 0.81176(30) &-0.2010(54) &   19.19 &
   0.81233(30) &-0.2048(54) &    8.06 &
   0.81290(30) &-0.2085(54) &    5.08
\end{tabular}
\end{table}

%
%
\begin{table}
\caption{
         Estimates of the critical exponent $\nu$ obtained from the
         slope of the
         cumulants at $K_c=0.81563(3)$ using eq.(\protect\ref{nulat}).
         }
         \label{nutab}
\begin{tabular}{clll}
\multicolumn{4}{c}{Estimates of $\nu$  from the data of $U_{L}$}\\ \hline
$L_1-L_2$ & $K_c-\Delta K_c$ & $K_c$ & $K_c+\Delta K_c$ \\ \hline
 4- 8  &0.6330(58) &0.6332(58) &0.6334(58) \\
 8-16  &0.6543(67) &0.6548(67) &0.6552(67) \\
16-32  &0.6617(83) &0.6628(83) &0.6639(83) \\
32-64  &0.6584(99) &0.6615(99) &0.6646(99) \\
 \hline\hline
\multicolumn{4}{c}{Estimates of $\nu$  from the data of $U_{L/2}$}\\ \hline
$L_1-L_2$ & $K_c-\Delta K_c$ & $K_c$ & $K_c+\Delta K_c$ \\ \hline
 4- 8  &0.5628(27) &0.5628(27) &0.5629(27) \\
 8-16  &0.6391(40) &0.6393(40) &0.6395(40) \\
16-32  &0.6540(52) &0.6546(52) &0.6552(52) \\
32-64  &0.6601(65) &0.6617(64) &0.6634(64) \\
  \hline\hline
\multicolumn{4}{c}{Estimates of $\nu$  from the data of $N\!N$}\\ \hline
$L_1-L_2$ & $K_c-\Delta K_c$ & $K_c$ & $K_c+\Delta K_c$ \\ \hline
 4- 8  &0.8095(57) &0.8097(57) &0.8099(57) \\
 8-16  &0.7451(49) &0.7456(49) &0.7460(49) \\
16-32  &0.6963(44) &0.6974(44) &0.6984(44) \\
32-64  &0.6819(52) &0.6847(52) &0.6877(52)
\end{tabular}
\end{table}

%
\begin{table}
\caption{
         Estimates of the critical exponent $\nu$ obtained from the
         fit of cumulants following eq.(\protect\ref{slope})
         at $K_c=0.81563(3)$. \# gives the number of discarded
         data-points with small $L$.
        }
         \label{nufittab}
\begin{tabular}{cccccccccc}
 \multicolumn{10}{c}{Estimates of $\nu$ from the data of  $U_L$}\\ \hline
&\multicolumn{3}{c}{ $K_c-\Delta K_c$}&
 \multicolumn{3}{c}{ $K_c$}&
 \multicolumn{3}{c}{ $K_c+\Delta K_c$}\\ \hline
\#    & $\nu$      & $C_{\nu}$ &$\G$& $\nu$& $C_{\nu}$
& $\G$  & $\nu$ & $C_{\nu}$ & $\G$  \\ \hline
0& 0.6513(18) & 0.0505(5)  &  4.49 &
   0.6522(18) & 0.0507(5)  &  5.09 &
   0.6532(17) & 0.0509(5)  &  5.79 \\
1& 0.6598(22) & 0.0536(8)  &  0.58 &
   0.6605(22) & 0.0538(8)  &  0.38 &
   0.6613(22) & 0.0540(8)  &  0.34 \\
2& 0.6635(34) & 0.0552(15) &  0.39 &
   0.6638(34) & 0.0553(15) &  0.10 &
   0.6642(34) & 0.0553(15) &  0.01 \\
\hline\hline
 \multicolumn{10}{c}{Estimates of $\nu$  from the data of $U_{L/2}$}\\ \hline
&\multicolumn{3}{c}{ $K_c-\Delta K_c$}&
 \multicolumn{3}{c}{ $K_c$}&
 \multicolumn{3}{c}{ $K_c+\Delta K_c$}\\ \hline
 \#  &   $\nu$    & $C_{\nu}$ & $\G$  & $\nu$
     & $C_{\nu}$ & $\G$  &
         $\nu$    & $C_{\nu}$ & $\G$  \\ \hline
0& 0.6216(9)  & 0.0386(2) &   186 &
   0.6222(9)  & 0.0387(2) &   190 &
   0.6229(9)  & 0.0389(2) &   195 \\
1& 0.6518(14) & 0.0490(5) &  4.90 &
   0.6522(14) & 0.0491(5) &  4.81 &
   0.6526(14) & 0.0492(5) &  4.86 \\
2& 0.6590(21) & 0.0521(9) &  0.61 &
   0.6592(21) & 0.0521(9) &  0.36 &
   0.6594(21) & 0.0522(9) &  0.25 \\
 \hline\hline
 \multicolumn{10}{c}{Estimates of $\nu$ from the data of $N\!N$}\\ \hline
&\multicolumn{3}{c}{ $K_c-\Delta K_c$}&
 \multicolumn{3}{c}{ $K_c$}&
 \multicolumn{3}{c}{ $K_c+\Delta K_c$}\\ \hline
 \# & $\nu$      & $C_{\nu}$ & $\G$  & $\nu$      & $C_{\nu}$
& $\G$  &
       $\nu$      & $C_{\nu}$ & $\G$  \\ \hline
0& 0.7347(13) & 0.1377(9)  &   166 &
   0.7355(13) & 0.1380(9)  &   161 &
   0.7363(13) & 0.1384(9)  &   156 \\
1& 0.7066(14) & 0.1160(11) & 31.39 &
   0.7073(14) & 0.1163(11) & 30.57 &
   0.7080(14) & 0.1166(11) & 30.07 \\
2& 0.6920(19) & 0.1038(15) & 1.48 &
   0.6924(19) & 0.1039(15) & 0.82 &
   0.6928(19) & 0.1040(15) & 0.69
\end{tabular}
\end{table}

%
%
\begin{table}
\caption{Estimates of the ratio  $\gamma/\nu$ obtained from the
         susceptibility at $K_c=0.81563(3)$
        using equation (\protect\ref{phen}).}
\label{gamnu}
\begin{tabular}{clll}
\multicolumn{4}{c}{Estimates of $\gamma/\nu$  from the data of $\chi_{L}$}\\
\hline
$L_1-L_2$ & $K_c-\Delta K_c$ & $K_c$ & $K_c+\Delta K_c$ \\ \hline
 4- 8   & 2.0663(20) & 2.0667(20) & 2.0669(20) \\
 8-16   & 2.0119(23) & 2.0128(23) & 2.0136(23) \\
16-32   & 1.9878(24) & 1.9902(24) & 1.9927(24) \\
32-64   & 1.9660(25) & 1.9728(25) & 1.9796(25) \\
 \hline\hline
\multicolumn{4}{c}{Estimates of $\gamma/\nu$  from the data of $\chi_{L/2}$}\\
\hline
$L_1-L_2$ & $K_c-\Delta K_c$ & $K_c$ & $K_c+\Delta K_c$ \\ \hline
 4- 8   & 1.9035(14) & 1.9039(14) & 1.9041(14) \\
 8-16   & 1.9630(16) & 1.9635(16) & 1.9642(16) \\
16-32   & 1.9735(18) & 1.9754(18) & 1.9773(18) \\
32-64   & 1.9625(19) & 1.9677(19) & 1.9729(19)
\end{tabular}
\end{table}

%
\begin{table}
\caption{Estimates of the ratio $\gamma/\nu$ obtained from
         the fit of the susceptibility following
         eq.(\protect\ref{chifinite})
         at $K_c=0.81563(3)$. \# gives the
         number of discarded data-points with small $L$.}
\label{gamnufit}
\begin{tabular}{cccccccccc}
 \multicolumn{10}{c}{Estimates of $\gamma/\nu$ from the data of  $\chi_L$}\\
\hline
&\multicolumn{3}{c}{ $K_c-\Delta K_c$}&
 \multicolumn{3}{c}{ $K_c$}&
 \multicolumn{3}{c}{ $K_c+\Delta K_c$}\\ \hline
\#   &$\gamma/\nu$& $C_{\gamma/\nu}$ & $\G$  &$\gamma/\nu$ & $C_{\gamma/\nu}$ &
 $\G$ &$\gamma/\nu$& $C_{\gamma/\nu}$ & $\G$  \\ \hline
0& 2.0107(5) & 1.0394(14) & 500 &
   2.0128(5) & 1.0356(14) & 448 &
   2.0149(5) & 1.0320(14) & 399 \\
1& 1.9910(6) & 1.1128(24) &  64.16 &
   1.9927(6) & 1.1091(24) &  46.56 &
   1.9943(6) & 1.1054(24) &  39.61 \\
2& 1.9835(10) & 1.1468(42) &  29.64 &
   1.9844(10) & 1.1452(42) &   9.48 &
   1.9854(10) & 1.1436(42) &   3.74 \\
 \hline\hline
 \multicolumn{10}{c}{Estimates for $\gamma/\nu$ from the data of $\chi_{L/2}$}\\
\hline
&\multicolumn{3}{c}{ $K_c-\Delta K_c$}&
 \multicolumn{3}{c}{ $K_c$}&
 \multicolumn{3}{c}{ $K_c+\Delta K_c$}\\ \hline
\#   &$\gamma/\nu$& $C_{\gamma/\nu}$ & $\G$  &$\gamma/\nu$ & $C_{\gamma/\nu}$ &
 $\G$ &$\gamma/\nu$& $C_{\gamma/\nu}$ & $\G$  \\ \hline
0& 1.9518(4) & 0.2022(2) & 524 &
   1.9535(4) & 0.2016(2) & 568 &
   1.9552(4) & 0.2010(2) & 620 \\
1& 1.9693(5) & 0.1911(3) & 28.52 &
   1.9706(5) & 0.1907(3) & 12.96 &
   1.9719(5) & 0.1902(3) &  8.53 \\
2& 1.9733(7) & 0.1885(5) & 30.10 &
   1.9741(7) & 0.1882(5) &  7.76 &
   1.9749(7) & 0.1880(5) &  0.74
\end{tabular}
\end{table}

\begin{table}
\caption{
         Comparison of critical properties of the 3D AF Potts model and
         the 3D $XY$ model determined by various methods.
        }
         \label{compare}
\begin{tabular}{llllll}
Method   & Ref.  & $K_c$  & $\gamma/\nu$ & $\nu$ & $U_L$ \\
\hline\hline
\hfill&\multicolumn{5}{c}{3D AF Potts model}\\ \hline
Phenomenological RG &  this work  & 0.81563(3) & 1.973(9) & 0.6639(38)
& 0.5862(6) \\
Phenomenological RG &\cite{wang}  & 0.8157(5)  & 1.99(3)  & 0.66(3)
& $-$        \\
Interface approach   &\cite{ueno}  & 0.810      & 1.10(2)  & 0.58(1)
& $-$        \\
MC                  &\cite{ono}   & 0.80       &   $-$    & $-$
& $-$        \\
MC                  &\cite{hoppe} & 0.781(25)  &  $-$     & $-$
& $-$        \\
High temperature series&\cite{yasumu}& 0.826      &  $-$     &  $-$
& $-$        \\ \hline
\hfill&\multicolumn{5}{c}{3D $XY$ model}\\ \hline
Phenomenological RG & \cite{wexy}& 0.45419(2) & 1.976(6) & 0.662(7)
& 0.5875(28)\\
High temperature MC & \cite{wexy}& 0.454170(7)&  $-$      & $-$
&   $-$     \\
Resummmed perturbation series&\cite{guillou80}& $-$ & 1.967(4)& 0.669(2)
& $-$       \\
$\epsilon$-expansion&\cite{brezin85a}& $-$ &  $-$ &  $-$ &
0.552 \\
Experiment $^4$He   &\cite{ahlers}& $-$     & $-$     & 0.6705(6)
& $-$  \\
\end{tabular}
\end{table}


\newpage
\baselineskip 4ex
\begin{thebibliography}{99}

\bibitem{Lehrbuch}
       See, for example:\\
       J.B. Kogut, Rev.Mod.Phys. {\bf 55}, 775 (1983).

\bibitem{KT}
       See, for example:\\
       J.M. Kosterlitz and D.J. Thouless, J.Phys. {\bf C6}, 1181 (1973);\\
       J.M. Kosterlitz, J.Phys. {\bf C7}, 1046 (1974);\\
       S.T. Chui and J.D. Weeks, Phys.Rev. {\bf B14}, 4978 (1976);\\
       J.V. Jos\'e, L.P. Kadanoff, S. Kirkpatrick and D.R. Nelson,
       Phys.Rev. {\bf B16}, 1217 (1977);\\
       T. Ohta and K. Kawasaki, Prog.Theor.Phys. {\bf 60}, 365 (1978).

\bibitem{banavar}
       J.R. Banavar, G.S. Grest, and D. Jasnow, Phys.Rev.Lett.
       {\bf 45}, 1424 (1980);\newline
       J.R. Banavar, G.S. Grest, and D. Jasnow,
       Phys.Rev. {\bf B25} , 4639 (1982).

\bibitem{hoppe}
       B. Hoppe and L.L. Hirst, J. Phys.  {\bf A18}, 3375 (1985);\newline
       B. Hoppe and L.L. Hirst, Phys.Rev. {\bf B34}, 6589 (1986).

\bibitem{ono}
       I. Ono, Prog.Theor.Phys.Suppl. {\bf 87}, 102 (1986).

\bibitem{ueno}
       Y. Ueno, G. Sun and I. Ono, J.Phys.Soc.Jpn. {\bf  58}, 1162 (1989).

\bibitem{wang}
       J.-S. Wang , R.H. Swendsen, and R. Koteck\'y,
       Phys.Rev.Lett. {\bf  63}, 109 (1989);\newline
       J.-S. Wang , R.H. Swendsen, and R. Koteck\'y,
       Phys.Rev. {\bf B42} , 2465 (1990).

\bibitem{wilson72}
       K.G. Wilson and M.E. Fisher, Phys. Rev. Lett. {\bf 28}, 240 (1972).

\bibitem{guillou80}
      J.C. Le Guillou and J. Zinn-Justin, Phys. Rev.  {\bf B21}, 3976 (1980).

\bibitem{guillou85a}
      J.C. Le Guillou and J. Zinn-Justin, J. Phys. Lett.  (Paris) {\bf 46},
       L137 (1985).

\bibitem{privman}
      V. Privman, P.C. Hohenberg, and A. Aharony, In {\it Phase Transitions and
      Critical Phenomena}, Vol 14, C. Domb and J.L. Lebowitz , eds.
      (Academic Press, London, 1991).

\bibitem{meyer} Preliminary results for the values of the cumulants at
      $K=0.8157$ on lattices up to $L=32$ by M.~Hasenbusch and S.~Meyer
      were reported by S.~Meyer in a seminar talk at DESY in February 1990.

\bibitem{binder}
      K. Binder, Phys.Rev.Lett. {\bf 47}, 693 (1981) ; \newline
      K. Binder, Z. Phys.  {\bf B43}, 119 (1981).

\bibitem{wolff89a}
      U. Wolff, Phys. Rev. Lett. {\bf 62},  361 (1989);\newline
      U. Wolff, Nucl. Phys. {\bf B322},  759 (1989).

\bibitem{swendferr}
      A.M. Ferrenberg and R.H. Swendsen, Phys. Rev. Lett. {\bf 61}, 2635 (1988).

\bibitem{siam}
      R.G. Miller, Biometrica {\bf 61}, 1 (1974); \newline
      B. Efron, {\it The Jackknife, the Bootstrap and other Resampling Plans}
      (SIAM, Philadelphia, PA, 1982).

\bibitem{wexy}
      A.P. Gottlob and M. Hasenbusch, Physica {\bf A201}, 593 (1993).

\bibitem{wegner}
      F.J. Wegner, Phys. Rev. {\bf B5}, 4529 (1972).

\bibitem{kadanoff}
      L.P. Kadanoff, In {\it Phase Transitions and Critical Phenomena} Vol. 5A,
      C. Domb and M.S. Green, eds. (Academic Press, London,1976).

\bibitem{janke90a}
      W. Janke, Phys. Lett. {\bf A148},  306 (1990).

\bibitem{yasumu}
      K. Yasumura and T. Oguchi, J. Phys. Soc. Jpn. {\bf 53}, 515 (1984).

\bibitem{ahlers} L.S. Goldner and G. Ahlers, Phys. Rev. {\bf B45}, 13129
      (1992).

\bibitem{landau} A.M. Ferrenberg and D.P. Landau, Phys. Rev. {\bf B44}, 5081
      (1991).

\bibitem{jankeo3}
      C. Holm and W. Janke, Phys.Lett {\bf A173}, 8 (1993).

\bibitem{brezin85a}
      E. Br\'{e}zin and J. Zinn-Justin, Nucl. Phys. {\bf B257}, 867 (1985).

\bibitem{hasenbus} M. Hasenbusch, Physica  {\bf A197}, 423 (1993).

\end{thebibliography}
\end{document}